\documentstyle[aps,twocolumn,epsfig]{revtex}

\begin{document}


\wideabs{
\title{Complexity in dislocation dynamics: model}

\author{M.-Carmen Miguel$^{1}$\thanks{Present address:
Departament de Fisica Fonamental, Facultat de Fisica, Universitat de
Barcelona, Av. Diagonal 647, 08028 Barcelona,
Spain. E-mail: carmen@ffn.ub.es}, Alessandro Vespignani$^1$, Stefano
Zapperi$^2$\\ J\'er\^ome Weiss$^{3}$, and Jean-Robert Grasso$^{4}$}

\address{$^1$The Abdus Salam International Centre for Theoretical Physics,
P.O. Box 586, 34100 Trieste, Italy\\ $^2$INFM, Universit\`a "La
Sapienza", P.le A. Moro 2, 00185 Roma, Italy\\ $^3$LGGE-CNRS, 54 rue
Moli\'ere, B.P. 96, 38402 St Martin d'H\'eres Cedex, France\\
$^4$LGIT, B.P. 53X, 38041 Grenoble Cedex 9, France}

\maketitle

\begin{abstract}
\noindent We propose a numerical model to study the viscoplastic
deformation of ice single crystals. We consider long-range elastic
interactions among dislocations, the possibility of mutual
annihilation, and a multiplication mechanism representing the
activation of Frank-Read sources due to dislocation pinning.  The
overdamped equations of motion for a collection of dislocations are
integrated numerically using different externally applied
stresses. Using this approach we analyze the avalanche-like
rearrangements of dislocations during the dynamic evolution.  We
observe a power law distribution of avalanche sizes which we compare
with acoustic emission experiments in ice single crystals under creep
deformation.  We emphasize the connections of our model with
non-equilibrium phase transitions and critical phenomena.

\end{abstract}
}

\section{INTRODUCTION}

The viscoplastic deformation of crystalline materials, such as ice
single crystals, involves the motion of a large number of
dislocations. Dislocations may be incorporated into a crystal in its
growth process, or under deformation conditions. They can penetrate
into the material from the sample surfaces, or be generated by various
mechanisms, as for example, in what is usually called a Frank-Read
source, activated by the pinning of a dislocation
loop~\cite{Hirth92,Nabarro87}. The elastic force between a pair of
dislocation lines decreases algebraically with the interline distance;
it can be attractive or repulsive, depending on the orientation of
their respective Burgers vectors; and eventually a pair of
dislocations may annihilate at relatively short distances. As a
consequence of all of these features, the collective behavior of a
large number of these defects appears to be an amazingly rich problem,
suitable to be studied from very different points of view.

A new insight into the classical issue of viscoplasticity was recently
opened by Weiss and Grasso~\cite{Weiss97,Weiss00a,Weiss00b} after
their experimental study of the acoustic activity of ice crystals
under creep. The complex character of the collective dislocation
dynamics reveals itself in experiments of acoustic emission (AE), as
reported in our companion paper~\cite{Weiss00b} in this volume. As a
matter of fact, when a material is deformed under constant load (creep
experiment) and dislocation motion is the dominant mechanism for
viscoplastic deformation, a constant strain-rate regime usually
follows after the initial transient stage. Orowan's relation $\dot
\gamma = \rho_m b v$ is known to prevail under such conditions, where
$\gamma$ is the strain of the sample, $\rho_m$ is the density of
mobile dislocations, $b$ is the Burgers' vector, and $v$ is the mean
velocity of the dislocations. Obviously, this is a mean-field relation
which neglects temporal and spatial fluctuations of both the density
and the velocity fields. As a result of their interactions, however,
dislocations tend to move cooperatively giving rise to a rather
complex and heterogeneous slip process. Dislocations move in groups to
form slip bands. Moving dislocations can pile-up against stable
dislocation configurations such as walls or boundaries, which may
eventually break apart.

Given the amplitude threshold and the frequency range accessible to the
experimental apparatus, the AE signals detected seem to correspond to
the synchronous motion of several dislocations, likely to occur during
the breakaway of a pile of these defects, or the activation of a
multiplication source. Various measurements of the acoustic activity
recorded during a stress-constant step show that the AE signal takes
place in the form of bursts distributed according to a power law. This
behavior is certainly a consequence of the collective motion of
dislocations which spontaneously gives rise to an avalanche-like
dynamics, typical of slowly driven dissipative systems. The power law
distribution provides evidence of scale-free cooperative behavior
whose origin could be ascribed to nonequilibrium continuous phase
transitions~\cite{transitions} or self-organized
criticality~\cite{soc}.

The AE experiments, however, have only access to information resulting
from the interplay of various magnitudes. Thus, the physical
interpretation of the generated AE waves remains a major difficulty,
and constitutes the main motivation of this work.

\section{DESCRIPTION OF THE MODEL}

Our goal is to characterize the viscoplastic deformation of ice or a
similar crystalline material from the perspective of nonequilibrium
statistical mechanics. To do so, we propose a simplified model to
study the dynamics of a collection of dislocations.

Ice crystals deform essentially by slip on the basal plane $(0001)$
($xy$ plane), i.e. the motion of dislocation lines or loops takes
place by gliding on the $xy$ planes. The simplest Burgers vectors
${\bf b}$ of a dislocation in a crystal of hexagonal ice are the
primitive cell vectors of the conventional hexagonal lattice. We start
studying a two-dimensional (2d) model representing a cross section of
the crystal which is perpendicular to the basal planes and parallel to
one of these lattice vectors, that is, for example, the $xz$ plane. In
this way, the dislocations constrained to move in this plane have all
Burgers vectors parallel to the chosen lattice vector ${\bf
b}=(b,0,0)$ and move along fixed lines parallel to the $x$ axis.  We
also consider that all $N$ dislocations are of edge type, and that, on
average over many realizations, the number of dislocations with
positive and negative Burgers vectors is the same.

Several $2d$-models containing similar basic ingredients have been
proposed in the literature in the last few
years~\cite{Kubin87,Amodeo90,Groma93,Fournet96,Miguel99}. A basic
feature common to most models, is that dislocations interact with each
other through the long-range elastic stress field they produce in the
host material. An edge dislocation with Burgers vector $b$ located at
the origin gives rise to a shear stress $\sigma^{s}$ at a point
$(x,z)$ of the form

\begin{equation}\label{eq:1}
\sigma^{s}=bD\frac{x(x^2-z^2)}{(x^2+z^2)^2},
\end{equation} 

\noindent where $D=\mu/2\pi(1-\sigma)$ is a coefficient involving the
shear modulus $\mu$ and the Poisson's ration $\sigma$ for the
material.  In our model, we further assume that Peierls stress is zero
and that the dislocation velocities are linearly proportional to the
local stress. Experimental evidence supports such a relationship for
low stress conditions, which is indeed the case in our
model. Accordingly, the velocity of the $n$th dislocation, if an
external shear stress $\sigma^{e}$ is also applied, is given by

\begin{equation}\label{eq:2}
v_n=b_n(\sum_{m\neq n} \sigma^{s}_{nm} - \sigma^{e}).
\end{equation} 

As the number of dislocations in any real crystal exceeds by far the
number of defects we can handle in a computer, one usually introduces
periodic boundary conditions (PBC) to effectively extend the size of
our system. To avoid the discontinuities arising from truncating
long-range elastic interactions (see Eq.~\ref{eq:1}), we resort to the
Ewald summation method. In this way, we have exactly accounted for the
interaction of a dislocation with all the infinite periodic replicas
of another dislocation in a finite cell of dimensions $L\times
L$. Contrary to what is stated in Ref.~\cite{Groma93}, we do not find
any spurious results coming from the implementation of PBC in this
fashion.

When the distance between two dislocations is of the order of a few
Burgers vectors, the high stress and strain conditions close to the
dislocation core invalidate the results obtained from a linear
elasticity theory (i.e. Eq~.(\ref{eq:1})). In these instances,
phenomenological nonlinear reactions describe more accurately the real
behavior of dislocations in a crystal. In particular in our model, we
account for the {\em annihilation} of dislocations with opposite
Burgers vectors when the distance between them is shorter than
$2b$. Thus the core of one dislocation in our model has a radius
of size $b$. 

Another important feature of any computer model is the implementation
of a mechanism for the {\em multiplication} of dislocations in the
sample. It is widely believed that the Frank-Read
mechanism~\cite{Hirth92} is the most relevant for a gliding process of
dislocations under creep deformation. Indeed Frank-Read sources (FRS)
have been observed in ice. In a FRS multiplication occurs by pinning
of a dislocation segment on the basal planes due, for example, to a
defect in the crystal, or to the very presence of dislocation dipoles,
piles, and walls. If the local stress concentration is less than a
critical value, the pinned segment bows out by glide. Beyond this
critical value, the dislocation segment wraps around itself, creating
a new dislocation loop and restoring the original configuration. Thus
a sequence of loops forms continuously from the source until the local
shear stress drops below the activation value.  In order to simulate
this mechanism, we split our system of size $L$ into smaller cells
where, at each time step, we keep track of both the local fraction of
pinned dislocations and the stress. Pairs of dislocations of opposite
Burgers vectors are generated in a given cell if (i) there are pinned
particles and (ii) if the local stress is large compared to a
threshold value (the pair creation probability is proportional to the
local stress). Each pair of dislocations reduces the local value of
the stress, and we keep on generating pairs until the local stress
drops below the threshold value. In such a way, we simulate the
procession of dislocations often seen experimentally when a FRS is
activated in a material.

To follow the time evolution of $N$ dislocations, we integrate
numerically the $N$ coupled equations $dx_n/dt=b_n(\sum_{m\neq n}
\sigma^{s}_{nm} - \sigma^{e})$, where $\sigma^{s}_{nm}$ is given by
Eq.~\ref{eq:1}, using an {\em adaptive step size fifth-order
Runge-Kutta algorithm}.  Simulations start from a configuration of $N$
point-dislocations randomly placed on a square cell of size $L$. So
far, we have considered two different box sizes $L=200b$, and
$L=300b$, with an initial number of dislocations $N_0=800$, and
$N_0=1500$, respectively. We let the system relax in the absence of
external stress, until it reaches a metastable
arrangement. Annihilation and multiplication processes imply that the
number of dislocations $N$ in our system is not fixed in the course of
time. After the system has reached a metastable arrangement, the
volume fraction of dislocations $\phi=N \pi b^2/(Lb)^2$ ranges between
$1-5$\%. Once in these conditions, we apply a small external shear
stress and keep track of the various quantities describing the
dynamics of the dislocations. Averages are made over at least $50$
realizations starting with different random initial conditions.

\section{CREEP DYNAMICS}

In the evolution of our model, dislocations build up complex and
highly fluctuating patterns containing, for example, dislocation
dipoles and walls, similar to those experimentally
observed~\cite{Hahner98,Bako99}.

\begin{figure}
\centerline{\epsfig{file=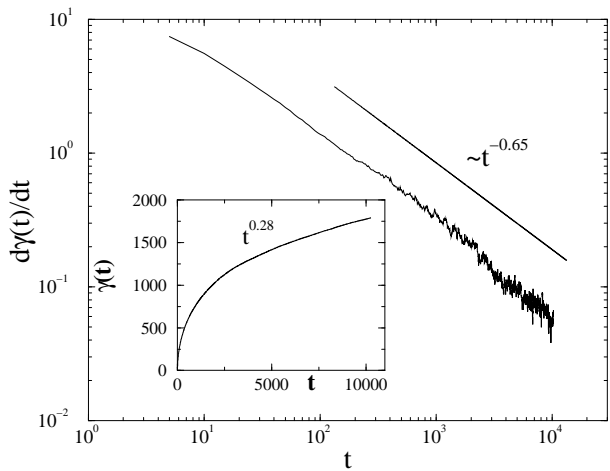,width=8truecm}} 
\caption{\small Mean strain rate as a function of time in our model of
collective dislocation dynamics. The curve shown in the inset
represents the global strain.  \label{sr}}
\end{figure}

If the external stress applied is not high enough to activate a
FRS~\cite{stress}, our system simply relaxes very slowly. After a
short initial transient, the root-mean-square velocity $(\sum_i
v_i^2)^{1/2}$ and the mean strain rate $d\gamma/dt \sim \sum_i b_i
v_i$ enter into a regime of power-law relaxation. In Fig.~\ref{sr} we
show in a double logarithmic scale the relaxation of $d\gamma/dt$. The
power-law exponent is around $-0.65$. This scale-free behavior seems
to correspond to what in the literature is known as {\em Andrade's
power law creep}, where the strain rate decays as $t^{-2/3}$, and,
consequently, the strain grows like $t^{1/3}$. The inset of
Fig.~\ref{sr} shows the strain curve. Our best fit is obtained with an
exponent of $0.28$.

One can further investigate the behavior of other quantities involved
in the collective dislocation dynamics, such as the moving dislocation
density, the annihilation rate, as well as the trajectories of the
dislocations. These results will be presented in detail
elsewhere~\cite{Miguel99}.

\begin{figure}
\centerline{\epsfig{file=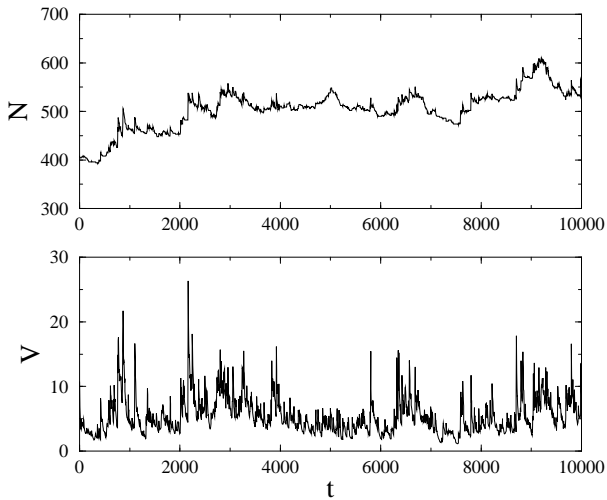,width=8truecm}} 
\caption{\small Number of dislocations $N$ and mean velocity $V$ as a
function of time in a given run of our model with
generation of dislocation pairs in FRS's.
\label{signal}}
\end{figure}

If we increase the external applied stress, the system starts to
generate dislocation pairs through the mechanism described in the
previous section. At this point, the system crosses over to a regime
in which the strain rate relaxes to a constant value, i.e. the {\em
linear creep regime}. In this regime, we can monitor observables such
as the average number of dislocations and the mean velocity
$V=\sum^{\prime} |v_i|$ of all moving dislocations $N_m$. In
Figure~\ref{signal}, we see that $V$ exhibits an intermittent and
bursty behavior that signals the presence of collective dislocation
rearrangements. Noticeably, the highest activity is not necessarily in
a one-to-one correspondence with the total number of dislocations $N$
present in the system.

\begin{figure}
\centerline{\epsfig{file=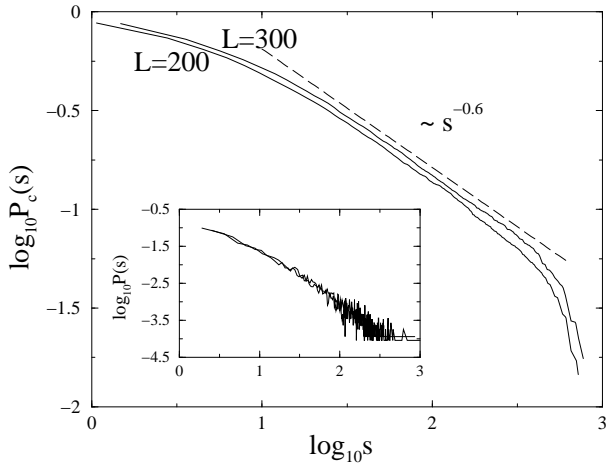,width=8truecm}} 
\caption{\small The cumulative {\em avalanche} size distribution $P_c$
for two system sizes $L=200,300$ represented in a double logarithmic
scale. The curves in the inset represent the raw distributions $P(s)$.
\label{distriS}}
\end{figure}

The features of this signal resemble those observed in apparently very
different systems, such as fracturing of wood and
concrete~\cite{frac}, Barkhausen effect~\cite{durin94}, and flux lines
in high-$T_c$ superconductors~\cite{nori95}, and induces us to think
in terms of avalanches. To define what an avalanche is in our system,
we introduce an offset value of the signal to eliminate what we call
the background noise. The latter can be thought of as the lower
experimental resolution, but at the same time corresponds to the
incoherent motion of $N_m$ independent dislocations. Thus, when the
signal exceeds this offset, we will have an avalanche. The avalanche
duration $T$ and magnitude $s=\sum^{T} V$ increase until the signal
drops down below the offset value. The cumulative distribution
$P_c(s)$ of the avalanches obtained after averaging over several
realizations is depicted in Figure~\ref{distriS} for the two system
sizes studied $L=200,300$.  We recover a clear power law distribution
($P_c(s)\sim s^{-\tau}$) extending over close to two decades. The
exponent $\tau\simeq 0.6$ is in reasonable agreement with experimental
data~\cite{Weiss97}. The universal properties of the distribution does
not depend on the particular choice of the offset value. The
distribution cut-off for large values of $s$ is due to the finite size
of the sample. (As one would expect, the cut-off is scaling
accordingly to the system size.)  This clearly points out the presence
of a very large (or infinite) characteristic size for the acoustic
events. It is worth remarking that larger stresses~\cite{stress}
introduce a characteristic scale in the process. A more detailed study
as a function of the applied stress is in progress~\cite{Miguel99}.

\section{CONCLUSIONS AND DISCUSSION}

The numerical investigation of the present model provides striking
evidence for the collective critical behavior of dislocation motion
under external stress. We have found two different regimes depending
on the externally applied stress: i) For low stresses, the mean
dislocation velocity and the mean strain-rate relax according to
Andrade's power law creep. ii) For higher stress values, the system
activates FRS's to which it responds exhibiting singular behavior in
the guise of avalanches distributed over many length scales.  The
avalanche distribution is a power law which signals the absence of any
characteristic size in the process.  Avalanche dynamics is the rule
rather the exception in slowly driven disordered
systems~\cite{frac,durin94,nori95}. Under the external drive (the
stress in the present case), the system jumps between metastable or
pinned configurations in which the dynamics is virtually frozen. In
the limit of a very slow driving the disordered energy landscape is
explored quasistatically and the response function exhibits critical
properties~\cite{ron99}.  Typically, a basic ingredient for this
behavior is the presence of quenched disorder acting as the source of
pinning in the system.  Noticeably, the system under study does not
contain any external source of disorder. Pinned states are due to the
various structures such as dipoles, piles, and dislocation walls, that
play the role of self-generated pinning centers that create the
pinning force landscape.  The new scenario poses many new and
interesting questions for a definitive identification and
understanding of the critical nature of dislocation dynamics.

\acknowledgements

We gratefully acknowledge M. Zaiser and R. Pastor-Satorras for fruitful
discussions and suggestions.


\begin{references}

\bibitem{Hirth92} J.P. Hirth and J. Lothe, 
{\it Theory of Dislocations} (Krieger Publishing Company, 1992).

\bibitem{Nabarro87} F.R.N. Nabarro, 
{\it Theory of Crystal Dislocations} (Dover, New York, 1992).

\bibitem{Weiss97} J. Weiss and J.R. Grasso, J. Phys. Chem. B 
{\bf 101}, 6113 (1997).

\bibitem{Weiss00a} J. Weiss, F. Lahaie, and J.R. Grasso, J. Geophys. Res.
{\bf 105}, 433 (2000).

\bibitem{Weiss00b} J. Weiss, J.R. Grasso, M.-Carmen Miguel,
A. Vespignani, and S. Zapperi, this issue (2000).

\bibitem{transitions} {\it Phase Transitions and Critical Phenomena},
edited by C. Domb and M.S. Green (Academic Press, London 1972-1976),
Vols. 1-17. 

\bibitem{soc} P. Bak, C. Tang, and K. Wiesenfeld, 
Phys. Rev. Lett. {\bf 59}, 381 (1987);
A. Vespignani and S.Zapperi, Phys. Rev. E, {\bf 57}, 6345 (1998).

\bibitem{Kubin87} J. Lepinoux and L.P. Kubin, Scripta Metall. 
{\bf 21}, 833 (1987).

\bibitem{Amodeo90} R.J. Amodeo and N.M. Ghoniem, Phys. Rev. B 
{\bf 41}, 6958 and 6968 (1990).

\bibitem{Groma93} I. Groma and G.S. Pawley, Phil. Mag. A, 
{\bf 67}, 1459 (1993).

\bibitem{Fournet96} R. Fournet and J.M. Salazar, Phys. Rev. B 
{\bf 53}, 6283 (1996).

\bibitem{Miguel99} M.C. Miguel, A. Vespignani, and S. Zapperi, in
preparation.

\bibitem{Hahner98} P. Hahner, K. Bay, and M. Zaiser, Phys. Rev. Lett. 
{\bf 81}, 2470 (1998).

\bibitem{Bako99} B. Bak\'o and I. Groma, Phys. Rev. B
{\bf 60}, 122 (1999).


\bibitem{stress} The stress values considered in our model are
comparable to the internal elastic stresses resulting from the
dynamics of our model.

\bibitem{frac}
A. Garcimartin, A. Guarino, L. Bellon, and S. Ciliberto,
Phys. Rev. Lett. {\bf 79}, 3202 (1997);
A. Petri, G. Paparo, A. Vespignani, A. Alippi, and M. Costantini
Phys. Rev. Lett. {\bf 73}, 3423 (1994).

\bibitem{durin94}
G. Bertotti, G. Durin, and A. Magni, 
J. Appl. Phys. {\bf 75}, 5490 (1994).

\bibitem{nori95}
S. Field, J. Witt, F. Nori and X. Ling. 
Phys. Rev. Lett. {\bf 74}, 1206 (1995).

\bibitem{ron99}
R. Dickman, M. A. Mu\~noz, A. Vespignani and S. Zapperi,
Braz. J. Phys. {\bf 30}, 27 (2000). 

\end{references}
\end{document}